\documentstyle[aps,prl,epsfig,floats]{revtex}
\begin{document}
\draft
\twocolumn[
\hsize\textwidth\columnwidth\hsize\csname@twocolumnfalse\endcsname

\title{Textures and the shapes of domains in Langmuir monolayers}

\author{Kok-Kiong Loh and Joseph Rudnick}
\address{Department of Physics, UCLA \\
405 Hilgard Ave., Los Angeles, California 90095-1547}
\date{\today}
\maketitle
\begin{abstract}
Two-dimensional domains containing an $XY$-like order parameter
exhibit non-trivial internal structure and take on shapes controlled
by the configuration that the order parameter adopts.  The
textures exhibited by the order parameter in such domains are 
controlled by the interplay between bulk and surface contributions 
to the energy.  We report calculations of the internal texture
and the shape of such domains.  These calculations lead to the
determination of the equilibrium properties of two-dimensional
domains, such as those observed in Langmuir monolayers.  This allows
for the unambiguous exploration of the implications of experimental 
findings.
\end{abstract}
]

Monolayers of surfactants confined to the air/water interface have
been found to possess complex textures similar to those observed in
liquid crystals.  The textures are generally observed in ``tilted''
phases---that is, in phases in which the long axes of the molecules in
the film are not perpendicular to the water surface but are uniformly
tilted with respect to the normal.  The textures are the result of the
spontaneous organization of the molecular tilt azimuth on macroscopic
length scales.  They can be understood, at least qualitatively, in
terms of continuum elastic theories of smectic liquid
crystals\cite{smectics} or, more relevant to the present work, in
terms of the two-dimensional $XY$ model.

When the surfactants organize into domains of condensed tilted phases,
such as the $L_2$ phase, surrounded by an isotropic
phase\cite{review}, many striking textures have been observed.  Such
textures range from continuous boojums, textures with point and/or
line defects\cite{smectics}, to striped textures in spiral
domains\cite{striped}.  We will, however, focus our discussions on
the boojums in which the tilt azimuth varies continuously and appears
to radiate in some cases from a defect located at the edge of the
domain or from a ``virtual'' defect in the isotropic phase\cite{L&S}.
Domains containing a boojum texture are not circular.  Rather, they
reflect the various nonisotropic influences that control domain shape
in the presence of a nonhomogeneous background.  Among the features
seen are protrusions, at times sharp enough to be characterized as
``cusps'' \cite{RudBru}, and indentations, which give the domains a
heartlike, or ``cardioid'' appearance \cite{cardioid,cigar}.

While there are notable cases in which the full analytical solution
can be obtained \cite{RudBru}, the determination of the properties of
a domain in which such a texture has formed presents a calculational
challenge.  Energy minimization entails the simultaneous adjustment of
the orientations of surfactant molecules in the interior of the domain
and the shape of the domain's boundary.  Past work on the problem of
the shape of surfactant domains has relied on either the assumption of
a texture that is unaffected by variations in the boundary
\cite{GalFou}, or on the approximation of a nearly circular
domain\cite{RudBru,bubble}.  Neither assumption is necessarily close
to reality, and results obtained in both cases can be legitimately
called into question.

This Letter describes a successful evaluation of equilibrium
properties of a domain of surfactants.  The evaluation is based on the
numerical solution of the extremum equations for the energy of an
$XY$-like texture confined to a compact, but not necessarily circular,
domain.  We are able to explore the texture and the associated domain
shape that result from various forms of the boundary and bulk energy
of the two-dimensional system.

The starting point in the analysis is the Hamiltonian of the $XY$
order parameter, represented as the two-dimensional unit vector
$\hat{c}(x,y)$, which is parameterized in terms of the angle,
$\Theta(x,y)$ between $\hat{c}(x,y)$ and the $x$ axis.  The order
parameter field $\Theta(x, y)$ will be referred to as the texture.  The
Hamiltonian, $H[\Theta]=\int_\Omega{\cal H}_b dA + \oint_\Gamma
\sigma(\vartheta-\Theta)\,ds$, where ${\cal H}_b$ is given by
\begin{eqnarray}
{\cal H}_b =&&\frac{\kappa}{2}\left\{ |\nabla \Theta|^2
+\beta
\left[\left( -\Theta_x^2+\Theta_y^2\right)\cos
2\Theta\right.\right.\nonumber\\
&&\left.\left.\,\,\,\,\,\,\,\,\,\,\,\,\,\,\,\,\,\,\,\,\,\,
\;\;\,\,\,\,\,\,\,\,\,\,\,\,\,\,\,\;\;
-2\Theta_x\Theta_y\sin 2 \Theta \right]\right\},
\label{Hamil1}
\end{eqnarray}
$\Theta_x$ and $\Theta_y$ represent partial derivatives of
$\Theta$ with respect to $x$ and $y$, respectively.
The integral $\int_{\Omega}$ is over the domain's bulk, while
$\oint_{\Gamma}$ is an integral over the closed curve bounding the
domain.  The coefficient $\kappa$ is the ``mean'' Frank constant, the
average of the bend and splay moduli, while $\beta$ is proportional to the
difference between the two moduli.  Specifically, $2\kappa = K_s+K_b$
and $2\kappa\beta = K_s-K_b$ where $K_{b}/2$ multiplies $|\vec{\nabla}
\times \hat{c}|^{2}$ in the energy of the $XY$ order parameter and
$K_{s}/2$ multiplies $|\vec{\nabla}\cdot \hat{c}|^{2}$.  The constants
$K_{b}$ and $K_s$ are, respectively, the bend and splay modulus.  The
boundary energy, $\sigma \left(\vartheta - \Theta \right)$, will have
the general form
\begin{equation}
\sigma(\varphi) = a_0 + a_1 \cos \varphi + a_2 \cos 2 \varphi + \cdots
\label{sigmaexpansion}
\end{equation}
The angle $\vartheta$ in the argument of the boundary energy is the
angle between the unit normal to the curve bounding the domain and the
$x$ axis.  The fact that the harmonic expansion,
(\ref{sigmaexpansion}), of the boundary energy consists entirely of
cosine terms reflects the absence of ``chiral'' interactions between
the texture and the domain boundary.

The minimization of the energy leads to equations for the texture
$\Theta(x,y)$ and the bounding curve $\Gamma$.  $\Theta(x,y)$
satisfies both a bulk and a boundary extremum equation.  The bulk
equation is
\begin{eqnarray}
-\nabla^2\Theta +
\beta\left[\left(\Theta_{xx}-\Theta_{yy}\right)\cos 2 \Theta
+2\Theta_{xy}
\sin 2 \Theta  \right. &&\nonumber \\
\left.+ \left(-\Theta_x^2+\Theta_y^2\right)\sin
2\Theta+2\Theta_x\Theta_y
\cos 2 \Theta\right]&&= 0. \label{bulk}
\end{eqnarray}
The double-subscripted $\Theta$'s represent second partial derivatives
with respect to the relevant variables.  The extremum equation for
$\Theta(x,y)$ on the boundary is
\begin{eqnarray}
\kappa\Theta_n\left[1 - \beta  \cos 2 (\vartheta-\Theta
)\right]&&\nonumber\\
+\kappa\beta
\Theta_t\sin2(\vartheta-\Theta)-\sigma^{\prime}(\vartheta-\Theta)&&=
0, \label{surface}
\end{eqnarray}
where $\Theta_n$ and $\Theta_t$ are respectively the normal and
tangential derivatives. The extremum equation for the bounding curve
$\Gamma$ is
\begin{eqnarray}
{\cal H}_b-\sigma^{\prime}(\vartheta-\Theta)\Theta_n-\sigma^{\prime\prime}
(\vartheta-\Theta)\Theta_t  && \nonumber \\
+\left[\sigma(\vartheta-\Theta)+\sigma^{\prime\prime}(\vartheta-\Theta)\right]
\frac{d\vartheta}{ds} + \lambda&&=0.
\label{boundary}
\end{eqnarray}
The quantity $\lambda$ is a Lagrange multiplier that enforces the
condition of constant enclosed area.

We first solve numerically for the equilibrium $\Theta(x, y)$ using a
variational formulation of the finite element method\cite{FEM},
assuming an initial boundary $\Gamma^{(0)}$.  A mesh of triangles is
generated over $\Omega$ using an adaptive method that refines the
grids where necessary.  Functions are defined by their values on the
vertices of the triangles.  The value of a function elsewhere is
obtained by interpolation.  The system energy $H[\Theta]$ is
then a function of the values of $\Theta(x, y)$ at the vertices, ${\bf
\Theta}\equiv(\Theta_i)$, which can be determined by solving $\partial
H({\bf\Theta})/\partial \Theta_i=0$\cite{recipes}, where $i$ runs from 1 to
the number of vertices.  We label the texture determined at this point
${\bf\Theta}^{(0)}$.

Equation (\ref{boundary}), after being cast into a coordinate-dependent
form, turns out to be a second order differential equation for
$\Gamma$.  Substituting ${\bf\Theta}^{(0)}$ for the $\Theta(x, y)$ in
the equation, $\Gamma$ can be determined by the Runge Kutta method
\cite{recipes}.  The solution, which is labeled $\Gamma^{(1)}$, is in
turn utilized to determine a new texture ${\bf\Theta}^{(1)}$.  The
process is iterated until self-consistency is achieved.

Making use of the numerical scheme outlined above, we have
investigated the the texture and shape of the bounding curve under the
influence of variations in the boundary energy coefficient $a_2$ in
Eq.  (\ref{sigmaexpansion}) and the stiffness coefficient $\beta$.
Before presenting our results, we note that when $a_2=0$ and
$\beta=0$, the exact result \cite{RudBru} is given by a circular
boundary of radius R, with the boojum texture, {\em i.e.}, a defect
with winding number +2\cite{GalFou} located at a distance $R_B\equiv
R(1+\sqrt{1+\rho^2})/\rho$ from the center of the domain, where
$\rho\equiv Ra_1/\kappa$ is the normalized domain radius.  The
normalized domain radius $\rho$ will be used as a gauge of the domain
size throughout the discussion.  For cases in which the domains are
not circular, $R$ is the effective radius, in that $\pi R^2$ will be
the area of the domain.

\begin{figure}
\centerline{\epsfig{file=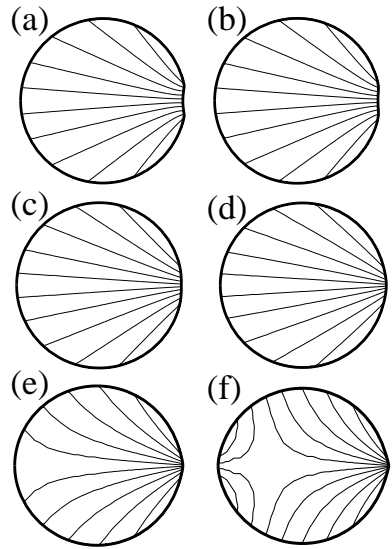, height=3in}}
\caption{The constant-order-parameter
contours and shapes of domains computed for $-0.5<a_2<0.5$,
$\rho=5$, $a_0=4$, $a_1=1$.  (a)$a_2=-0.5$, (b)$a_2=-0.3$, (c)
$a_2=-0.1$,( d) $a_2=0.1$,(e) $a_2=0.3$ and (f) $a_2=0.5$.}
\label{adep}
\end{figure}

\begin{figure}
\centerline{\epsfig{file=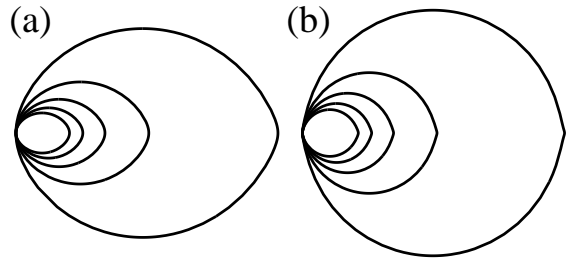, width=3in}}
\caption{The shapes of domains
of various sizes computed for $a_0=4$, $a_1=1$ and $a_2=0.6$.
(a)Smaller domains with $\rho=0.2, 0.25, 0.33, 0.5, 1$ which exhibit 2-fold
symmetry.  (b)Larger domains with $\rho=2, 2.5, 3.3, 5, 10$ which
have a protrusion on one end of the boundary.  Each of the sets of
domains are plotted to scale.}
\label{rdep}
\end{figure}

We first look at our result by varying only $a_2$ while keeping
$\beta=0$.  In the limit of very small domains, $\rho \ll 1$, the line
tension anisotropy has very little effect on the texture, which
exhibits almost no spatial variation.  The boundary response of the
domain is not affected by the $a_1$ contribution.  If $a_2\neq0$,
small domains exhibit twofold symmetry.  They become elongated when
$a_2>0$, and they flatten at both ends when $a_2 <0$.  For larger
domains in which $\rho \sim 1$ and $a_2/a_1\ll 1$, we find a
protrusion when $a_2 > 0$, and an indentation when $a_2 < 0$.  The
feature lies on the end of the domain boundary closest to the virtual
defect.  The loss of twofold symmetry is largely due to the fact the
dominant contribution of the domain texture, which in turn induces
deformation through $a_2$, comes from the $a_1$ contribution in the
boundary anisotropy.  Boundary features are more prominent as the
domain size increases.  Figure~\ref{adep} shows the shapes of domains
for $-0.5<a_2<0.5$ for $\rho=5$.  Numerically, the boundary features
are well-defined for domains of sizes $\rho \sim 1$.  To the numerical
accuracy that we are able to achieve, there is no discontinuity in
slope on the domain boundary.  That this ought to be the case can be
verified analytically\cite{futurework,erratum}.  The numerical result
we have is compatible with those discussed in Refs.
\cite{RudBru,GalFou,futurework}.  For even larger domains, $\rho\gg1$,
the features are confined in a small portion of the boundary.  The
domains become nearly circular again in the large-$\rho$ limit.  Up to
the largest domain we have examined, $\rho=32$, we can identify
protrusions when $a_2/a_0 > 0.1$.  Because of the rapid texture
variation in the immediate vicinity of the boundary, associated with
the approach to the boundary of the virtual boojum singularity, we are
unable to perform dependable numerical investigations of extremely
large domains.  This leaves open the question of the asymptotic
behavior of the domain in the large-$\rho$ regime.  Figure~\ref{rdep}
shows the shapes of domains in which the normalized radius ranges from
$\rho=0.2$ to $\rho=10$ for $a_1=1$ and $a_2=0.6$.

\begin{figure}
\centerline{\epsfig{file=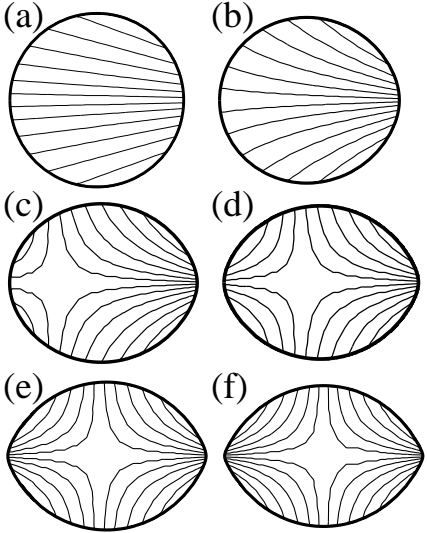, height=3in}}
\caption{The constant-order-parameter contours and the shapes of domains
with $\rho=1$ and $a_0=4$.  The coefficients of the anisotropic
line-tension are (a)$a_1=1$, $a_2=0$, (b)$a_1=0.8$, $a_2=0.2$,(c)$a_1=0.6$,
$a_2=0.4$,(d)$a_1=0.4$, $a_2=0.6$,(e)$a_1=0.2$, $a_2=0.8$ and
(f)$a_1=0$, $a_2=1$.}
\label{cigarshape}
\end{figure}

As regards the texture, we are able to numerically reproduce the
exact result of Rudnick and Bruinsma \cite{RudBru} when $a_1=0$ and
$a_2=1$ and the boundary is fixed at a circle.  This texture is
associated with two virtual defects.  When the boundary is allowed to
relax, the domain acquires a ``cigar shape''.  When both $a_{1}$ and
$a_2$ are not equal to zero, the texture can be thought of as a
superposition of pure $a_1$ and pure $a_2$ textures.  Typically for
$\rho\sim1$, the effect of the second defect becomes observable, in
the form of a distortion of constant-order-parameter contours, when
$a_2/a_1\approx1/4$.  The progressive changes of the texture and
domain shape from a system with pure $a_1$ to a system with pure $a_2$
are depicted in Fig.~\ref{cigarshape}.  Domains with indentations,
protrusions, and cigar-shaped domains, have all been observed
\cite{cigar}.

We now investigate the textures and boundaries that result when $\beta
\neq 0$.  The parameters $a_{0}$ and $a_{1}$ are assumed to be finite,
while all other $a_i$'s are set equal to zero.  We have computed the
textures and domain shapes for $-0.8<\beta<0.8$ and $0.5<\rho<8$.
When $|\beta|\leq 0.5$, our results are in qualitative agreement with
those reported in Ref.  \cite{GalFou}.  For $\beta < 0$, the texture
is modified as if the virtual singularity has moved closer to the
domain boundary, and the domain acquires a very small protruding
feature.  On the other hand, when $\beta > 0$, the texture relaxes as
if the virtual defect has moved away from the boundary, and the
boundary correction is similar to that of a domain with an
indentation.  There are, however, quantitative differences.  The
textures obtained numerically deviate significantly from the boojum
textures utilized in Refs.  \cite{GalFou,RivMeu}.  The sizes of the
features on the boundaries are no more than $1\%$ of the
overall domain radii.  We do not expect such features to be observable
experimentally.

\begin{figure}
\centerline{\epsfig{file=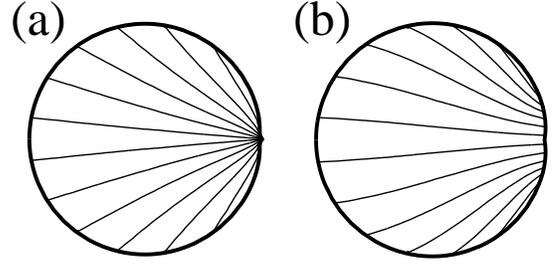, width=3in}}
\caption{The constant-order-parameter contours and the shapes of
domains with $\rho=8$, $a_0=4$, $a_1=1.6$ and $a_2=0$.
Their stiffness coefficients are (a)$\beta=-0.8$, (b)$\beta=0.8$.}
\label{bdep}
\end{figure}

\begin{figure}
\centerline{\epsfig{file=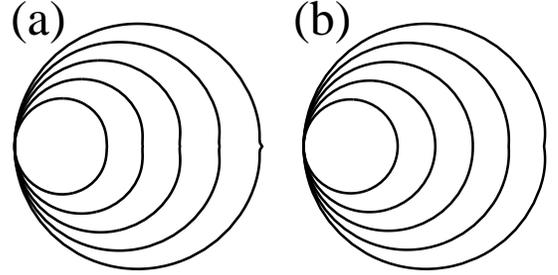, width=3in}}
\caption{The domain shapes computed for $a_0=4$, $a_1=1.6$ and $a_2=0$,
and $\rho=0.5, 1, 2, 4, 8$.
Their stiffness coefficients are (a)$\beta=-0.8$, (b)$\beta=0.8$.  For
ease of observation, domains are not shown to scale.}
\label{brdep}
\end{figure}

When $|\beta|=0.8$ and $\rho = 8$, the domains are no longer circular
in appearance.  Domains acquire small protrusions for negative $\beta$.
Domains have an indentation when $\beta$ is positive.  Figure~\ref{bdep}
shows the domain shapes and textures computed with $\beta=\pm0.8$ for
$\rho=8$.  The deviation from the boojum texture is more clearly
observable when $\beta = +0.8$ in Fig.~\ref{bdep}b.  The
constant-order-parameter contours distort so as to push the virtual
defect away from the boundary.  Figure~\ref{brdep} shows the changes in
the domain shape when $\rho$ decreases from $8$ to $0.5$ for
$\beta=\pm 0.8$.  The amplitude of the protrusion of the domain with
negative $\beta$ decreases with $\rho$.  At the same time, segments of
the boundary with negative curvature start to develop adjacent to the
protrusion.  At $\rho=2$, the negative curvature segments of the
boundary overwhelm the protrusion and give the domain an indentation.
When $\rho=0.5$, the indentation disappears, while the protrusion
continues to decrease in amplitude.  The domain appears circular.  The
complex boundary behavior described above can be attributed to the
rapid variation of the order parameter along the boundary near the
singularity.  In contrast to this behavior, the size of the
indentation on the boundary decreases monotonically as $\rho$
decreases when $\beta = 0.8$.  The domain appears circular at $\rho
\leq 2$.  It is the strength of the splay modulus that smooths out the
variation of the order parameter along the segment of the boundary
near the singularity and results in simple boundary behavior.

Our attempts to compare our results with experiments indicate that all
the observed domain shapes in Refs.  \cite{cigar,bubble,RivMeu} can be
accounted for, a least qualitatively, by line-tension anisotropy
alone.  Although we have established the existence of nontrivial
domain shapes due to purely elastic anisotropy, domains with the
protrusions and indentations discussed immediately above occur only
when the anisotropy is very strong $|\beta|=0.8$.  Furthermore, the
protrusions that are generated in our calculations do not resemble
those reported to have been observed experimentally \cite{bubble,RivMeu}.
We conclude that line-tension anisotropy must be present in such
monolayers.  On the other hand, both line-tension and elastic anisotropies
may be responsible for indentations on the boundary of domains of the
compressed phase.  Further measurements on the size dependence of the
domain shapes are needed to identify the underlying mechanism.

We note here that our analysis is based on a highly simplified model
of monolayers.  Other factors that contribute to the boundary shapes
and textures---such as the dipolar interaction \cite{dipolar}, the
tilt degree of freedom \cite{tilt} and the bond-orientational ordering
\cite{smectics}---have been neglected.  However, the results reported
here are useful in that they do indicate the influence of various
features that make up the model discussed.

In conclusion, we have devised a numerical scheme that enables us to
solve simultaneously for the minimum energy configuration of an $XY$
order parameter confined to a two-dimensional domain and the extremal
shape of the boundary of that domain.  We have utilized the method to
investigate the response of the texture-boundary system under the
influence of a nontrivial anisotropy in the boundary energy as well
as the bulk elastic energy.  Both line-tension anisotropy and elastic
anisotropy result in nontrivial domain shapes.  The observed domains
with protrusions and cigar shape can be attributed to the line-tension
anisotropy.  On the other hand, elastic anisotropy cannot be
ruled out as the source of the observed cardioids.

We are grateful to Professor Charles Knobler, Professor Robijn Bruinsma and
Dr.  Jiyu Fang for very useful discussions.


\begin{references}
\bibitem{erratum} In light of this result, the claims in
\cite{RudBru} of a cusp-like feature in the form of a discontinuity in
the slope of the bounding curve cannot be true.  In fact, the result
is the consequence of an error in the analysis reported in that paper.
\end{references}
\end{document}